\def\appropto{\mathrel{\vcenter{
  \offinterlineskip\halign{\hfil$##$\cr
    \propto\cr\noalign{\kern2pt}\sim\cr\noalign{\kern-2pt}}}}}

\documentclass[twocolumn]{aastex62}

\received{1 March 2019}
\revised{30 May 2019}
\accepted{5 June 2019}
\shorttitle{CGM Pressure Profiles}
\shortauthors{Voit et al.}

\begin{document}

\title{\bf Circumgalactic Pressure Profiles Indicate Precipitation-Limited Atmospheres for $M_* \sim 10^9$--$10^{11.5} \, M_\odot$}

\correspondingauthor{G.\ Mark Voit}
\email{voit@pa.msu.edu}

\author{G.\ Mark Voit$^1$, Megan Donahue}  
\affiliation{Department of Physics and Astronomy,
                 Michigan State University,
                 East Lansing, MI 48824} 
                 
\author{Fakhri Zahedy$^2$, Hsiao-Wen Chen}
\affiliation{Department of Astronomy \& Astrophysics,
                 The University of Chicago} 

\author{Jessica Werk}
\affiliation{Department of Astronomy,
                 University of Washington} 

\author{Greg L. Bryan}
\affiliation{Department of Astronomy,
                 Columbia University} 
\affiliation{Center for Computational Astrophysics,
                Flatiron Institute} 

\author{Brian W. O'Shea$^1$}

\begin{abstract}
Cosmic gas cycles in and out of galaxies, but outside of galaxies it is difficult to observe except for the absorption lines that circumgalactic clouds leave in the spectra of background quasars.  Using photoionization modeling of those lines to determine cloud pressures, we find that galaxies are surrounded by extended atmospheres that confine the clouds and have a radial pressure profile that depends on galaxy mass. Motivated by observations of the universe's most massive galaxies, we compare those pressure measurements with models predicting the critical pressure at which cooler clouds start to precipitate out of the hot atmosphere and rain toward the center. We find excellent agreement, implying that the precipitation limit applies to galaxies over a wide mass range.
\end{abstract}

%\vspace*{2em}


\section{Introduction}
\setcounter{footnote}{0}

Lyman Spitzer postulated in 1956 that a million-degree corona surrounds the Milky Way galaxy \citep{Spitzer_1956ApJ...124...20S}.  His proposal was based on absorption-line observations of much cooler gas clouds along lines of sight extending high above the galactic disk \citep{MunchZirin_1961ApJ...133...11M}. Hypothesizing that those clouds must be pressure-confined by a volume-filling ambient medium, Spitzer inferred its pressure from the pressures of the cooler clouds, and its temperature by assuming the ambient medium to be in hydrostatic equilibrium in the galaxy's potential well.  Direct confirmation of the corona's existence came only gradually, decades later, as X-ray observations began to distinguish its distinct contribution to the X-ray background that covers the sky \citep[e.g.,][]{KuntzSnowden_2000ApJ...543..195K,HenleyShelton_2013ApJ...773...92H}.  Astronomers of the twenty-first century refer to this gas as the circumgalactic medium (CGM) and have found that most of the baryons associated with a galaxy reside there, along with a large fraction of the elements produced by a galaxy's stars \citep[e.g.,][]{Chen_2010ApJ...714.1521C,Tumlinson_2011Sci...334..948T,Prochaska_2011ApJ...740...91P,Werk_2014ApJ...792....8W,Keeney_2017ApJS..230....6K,Zahedy_2019MNRAS.484.2257Z}.  

X-ray emission from the volume-filling hot component of the CGM around high-mass galaxies can be directly observed \citep[e.g.,][]{MathewsBrighenti2003ARAA..41..191M}, but it cannot yet be seen around galaxies with masses similar to or less than the Milky Way's \citep[e.g.,][]{Bregman_2007ARA&A..45..221B}.  Most of what we know about the CGM around those galaxies has therefore been inferred from observations of the absorption lines that it produces in the spectra of background quasars, which are most sensitive to $10^4$~K gas that is photoionized and heated by intergalactic ultraviolet (UV) radiation.  Those $10^4$~K clouds are thought to trace both the gaseous inflows that sustain star formation in galaxies and the galaxy-scale outflows that limit it \citep[e.g.,][]{TPW_CGM_2017ARA&A..55..389T}.  Such observations place strong constraints on models of the feedback processes that regulate galaxy evolution.  However, the observed velocity differences between a galaxy and the absorption-line clouds that surround it tend to be smaller than the expected speeds of Keplerian orbits \citep{Borthakur_2016ApJ...833..259B,Huang_2016MNRAS.455.1713H,McQuinnWerk_2018ApJ...852...33M,Zahedy_2019MNRAS.484.2257Z}, which makes it difficult to understand if those clouds are either falling ballistically toward the galaxy or being ejected at speeds sufficient to escape the galaxy's potential well.  

This Letter presents evidence indicating that the $10^4$~K clouds around galaxies are instead confined by a volume-filling corona that circulates the elements made by the galaxy's stars, but does not deviate far from hydrostatic equilibrium.  Those coronae appear to be similar to the X-ray--emitting ambient media around the most massive galaxies, in which energetic outflows powered by accretion of gas onto a central black hole balance radiative energy losses from the CGM.  In massive galaxies, the feedback loop connecting black hole accretion to the CGM suspends the ambient gas in a state in which the cooling time ($t_{\rm cool}$) required for the gas to radiate away its thermal energy cannot drop much below 10 times the freefall time ($t_{\rm ff}$) that it would take for a gas blob to fall freely to the center of the galaxy \citep{Sharma+2012MNRAS.427.1219S,Sharma_2012MNRAS.420.3174S,Gaspari+2012ApJ...746...94G,Voit_2015Natur.519..203V}.  Numerical simulations have shown that CGM gas with $t_{\rm cool} \approx 10 t_{\rm ff}$ tends to produce a rain of cold clouds that condense out of the ambient medium and fall toward the central black hole through a process known as ``chaotic cold accretion'' \citep{Gaspari+2013MNRAS.432.3401G,Gaspari_2017MNRAS.466..677G}.  The energy released as those clouds accrete onto the black hole then heats the ambient medium, causing it to expand, thereby increasing $t_{\rm cool}$ and reducing the precipitation of clouds.  An ambient CGM that is regulated through such a feedback loop to have $10 \lesssim t_{\rm cool}/t_{\rm ff} \lesssim 20$ is therefore ``precipitation-limited'' \citep{Voit_2017_BigPaper}, with important consequences for the rate at which condensation of CGM gas can fuel star formation \citep{Voit_PrecipReg_2015ApJ...808L..30V}.
 
 \section{CGM Pressure Measurements}
 
The evidence for a precipitation-limited CGM around less massive galaxies comes from photoionization models of the $10^4$~K clouds.  Those clouds are exposed to the ionizing UV background that permeates intergalactic space, which is known to better than a factor of 3 \citep[e.g.,][]{Shull_UVB_2015ApJ...811....3S}.  Each element in the cloud therefore has an ionization state depending on the ratio $U = n_\gamma / n_{\rm H}$ , where $n_\gamma$ is the number density of ionizing photons and $n_{\rm H}$ is the number density of hydrogen nuclei.  That ratio, known as the ionization parameter, is related to the cloud's pressure through
\begin{equation}
  - \log U = \log P - \log n_\gamma - \log T_{\rm eq}
\end{equation}
where $T_{\rm eq} \sim (0.5$--$2.0) \times 10^4$~K is the temperature at which photoelectric heating and radiative cooling balance each other and pressure has been expressed in terms of $n_{\rm H}T$.   For a given UV background, the quantity $- \log U$ is consequently a proxy for $\log P$.   

The {\em Hubble Space Telescope's} Cosmic Origins Spectrograph (COS) has obtained measurements of $U$ in $10^4$~K clouds around approximately 60 galaxies in the mass range $9 < \log M_* < 12$, where $M_*$ is the stellar mass (in solar units) inferred from a galaxy's luminosity and color.   This Letter analyzes a sample of CGM absorption-line clouds comprising six subsamples culled from three different studies. ÊOne of those studies was performed by the COS Guaranteed Time Observing (COS-GTO) team \citep{Stocke_2013ApJ...763..148S,Keeney_2017ApJS..230....6K}. ÊFrom that study, we took the 13 galaxies with absorption lines observed at $0.05 < r_{\rm proj} / r_{\rm halo} < 1$ around galaxies with $\log M_* > 9.0$, where $r_{\rm halo}$ is the dark-matter halo radius determined by those authors. ÊWe then subdivided them into three subsamples with $9.0 < \log M_* < 9.4$ (two galaxies), $9.4 < \log M_* < 9.8$ (two galaxies), and $10.2 < \log M_* < 11.3$ (nine galaxies). ÊAll the lower-mass galaxies have redshift $z < 0.1$, and all the higher-mass galaxies have $z < 0.2$. ÊThe second study was performed by the COS-Halos team \citep{Werk_2013ApJS..204...17W,Werk_2014ApJ...792....8W}. ÊFrom that study, we took the 32 galaxies with measured values of $U$ at $0.05 < r_{\rm proj} / r_{\rm halo} < 1$ and subdivided them into two subsamples with $9.5 < \log M_* < 10.3$ (11 galaxies) and $10.3 < \log M_* < 11.4$ (21 galaxies). ÊThose galaxies all fall into the redshift range $0.1 < z < 0.4$. ÊThe third study is COS-LRG \citep{Chen_2018MNRAS.479.2547C,Zahedy_2019MNRAS.484.2257Z}, which contains 11 luminous red galaxies (LRGs) with low-ionization CGM absorption lines strong enough to support measurement of $U$. ÊThey form a sixth subsample of galaxies that has $10.9 < \log M_* < 11.6$ and $0.2 < z < 0.6$.

Our COS-Halos and COS-LRG subsamples share three galaxies in common, along the lines of sight to quasars J0910+1014, J0950+4831, and J1550+4001. ÊThe COS-Halos study did not attempt to separate the absorption lines into distinct components and represent a fit to the entire column density derived for each ion used to determine $U$, whereas the COS-LRG team elected to perform separate ionization-parameter fits for components that can be distinguished from each other in velocity space. ÊIf there are multiple absorption-line clouds along a given line of sight through the CGM, then the simplest hypothesis for the pressure differences found among them is that those clouds are at different distances from the central galaxy within a confining atmosphere in which pressure declines with radius. ÊIf so, then the cloud with the greatest pressure would be most representative of the pressure at $r = r_{\rm proj}$. ÊOur study therefore includes only the COS-LRG points for the three galaxies jointly analyzed by COS-Halos and COS-LRG.  Even though this choice mitigates some of the potential projection effects, there may still be significant systematic uncertainties in the inferred values of $U$ resulting from complex, overlapping absorption lines. ÊIn the longer term, those uncertainties will need to be quantified and constrained with the use of physical models for those substructures \citep[e.g.,][]{Stern_2018arXiv180305446S}.

 % ----------------------------------------
\begin{figure}[t]
\begin{center}
\includegraphics[width=3.5in, trim = 0.1in 0.1in 0.0in 0.0in]{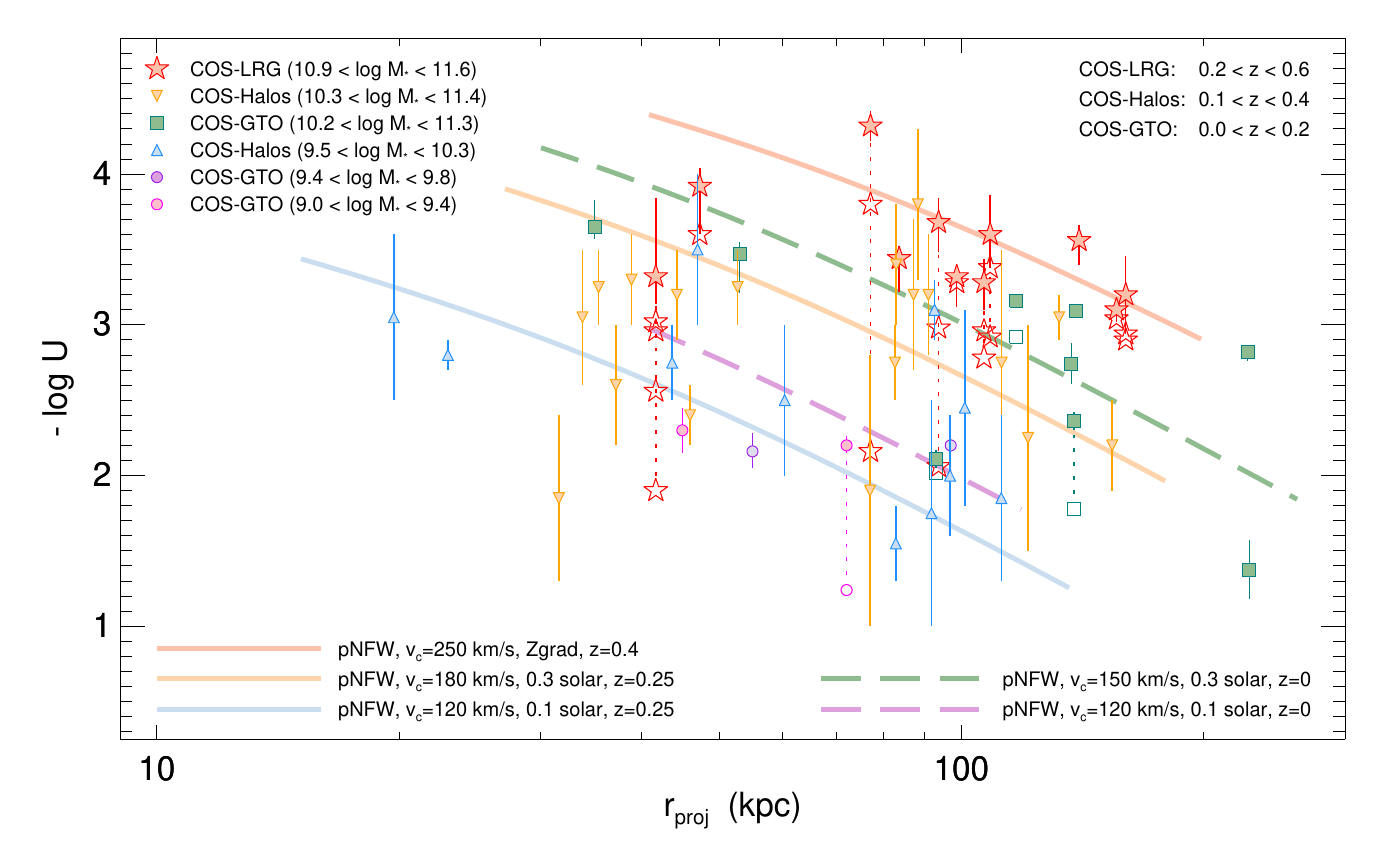} \\
\end{center}
\caption{ \footnotesize 
Dependence of circumgalactic cloud ionization level ($U$) on projected radius ($r_{\rm proj}$) among galaxies of differing stellar mass ($M_*$).  The vertical axis shows $- \log U$ because it is a proxy for cloud pressure.  Symbols show measurements of $U$ at $r_{\rm proj}$ for individual CGM clouds: triangles from \citet{Werk_2014ApJ...792....8W}, squares and circles from \citet{Keeney_2017ApJS..230....6K}, and stars from \citet{Zahedy_2019MNRAS.484.2257Z}.  The upper-left legend gives the range of $M_*$ corresponding to each symbol type.  Vertical dotted lines connect ionization measurements of multiple clouds along the same line of sight near a galaxy, and the open symbols represent clouds that have lower pressures than the highest-pressure cloud at the top of the line.  Comparison with the mass-segregated subsamples shows that galaxies of greater mass tend to have greater circumgalactic pressure.  Within each subsample, circumgalactic pressure declines with radius.  The thick solid and dashed lines represent models of precipitation-limited coronae from \citet{Voit_2018arXiv181104976V} for galaxies of approximately the same stellar mass and redshift ($z$) as those in the subsample with points of a corresponding color, as shown in the legend at the bottom.
\vspace*{1em}
\label{fig-U_vs_r}}
\end{figure}
% ----------------------------------------

% ----------------------------------------
\begin{figure*}[t]
\begin{center}
\includegraphics[width=6.0in, trim = 0.1in 0.1in 0.0in 0.0in]{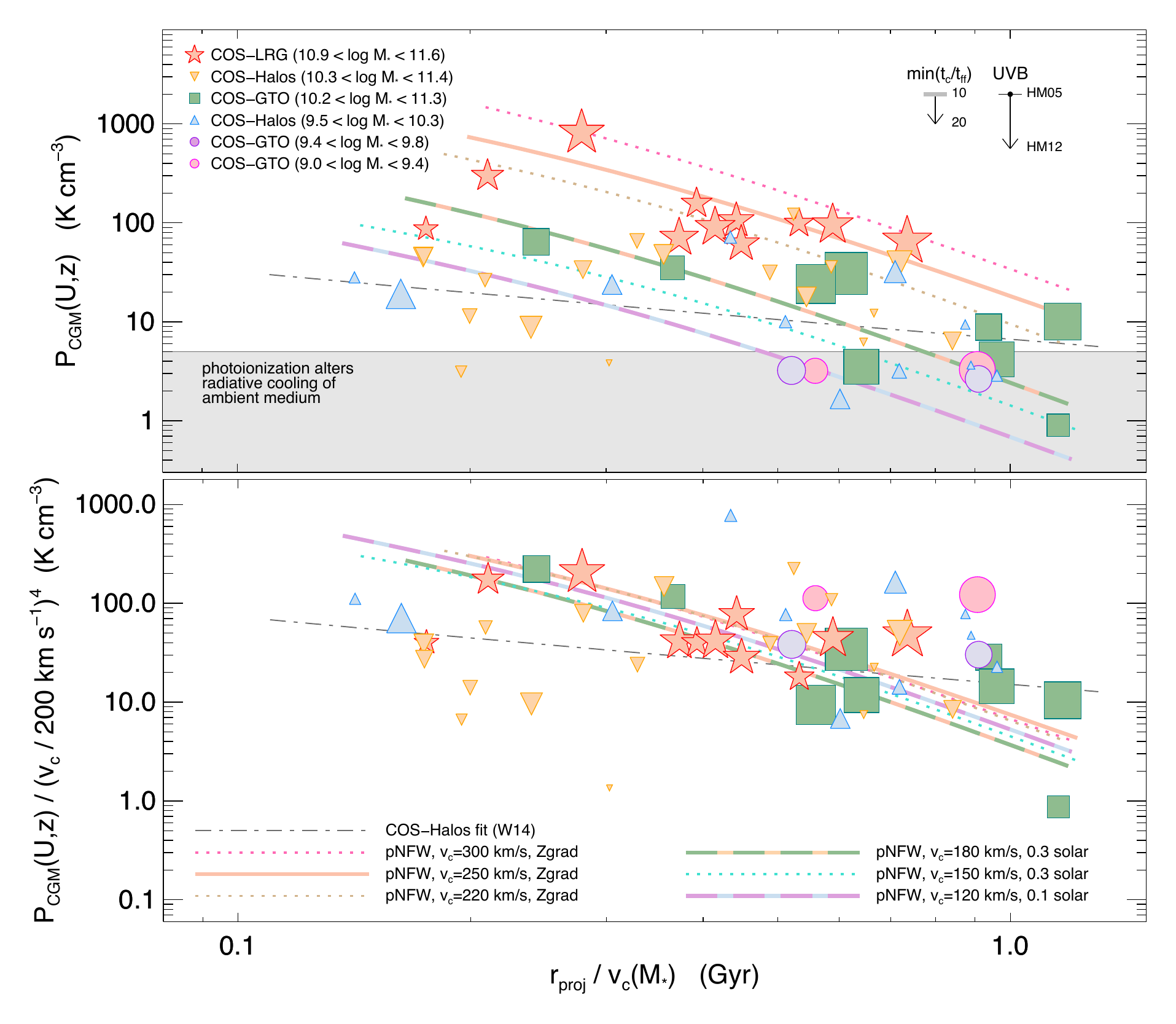} \\
\end{center}
\caption{ \footnotesize 
Dependence of circumgalactic pressure ($P_{\rm CGM}$) on the dynamical timescale obtained by dividing projected radius by the circular velocity ($v_c$) characterizing the potential well.  Symbol types represent the same subsamples as in Figure 1 and show pressures ($P_{\rm CGM}  = n_{\rm H}T$) inferred from observations of $U$, assuming the redshift-dependent HM05 UV background implemented in the photoionization code CLOUDY and based on \citet{HaardtMadau_2001cghr.confE..64H}.  Larger symbols have greater weight, which is proportional to the inverse square of uncertainty in $\log U$.  Colored solid and dashed lines have the same meanings as in Figure~\ref{fig-U_vs_r}, but with the redshift dependence removed.  Dotted lines show additional precipitation-limited models with intermediate $v_c$ values. The top panel shows that $P_{\rm CGM}$  at $r_{\rm proj} / v_c$ depends strongly on galaxy mass.  Downward arrows show how the model lines would shift if $\min ( t_{\rm cool} / t_{\rm ff})$ were assumed to be 20 instead of 10 and how the data points would shift if the UV background of \citet[][HM12]{HaardtMadau_2012ApJ...746..125H}  were assumed instead of HM05. The bottom panel shows that dividing $P_{\rm CGM}$  by $v_c^4$ greatly reduces the mass dependence among both the models and the data points.  Gray dotted-dashed lines show the power-law dependence of $U$ on $r_{\rm proj}$ originally determined by \citet{Werk_2014ApJ...792....8W} from a fit to the entire COS-Halos sample.  It is shallower than the trends in the individual subsamples, indicating that an analysis combining galaxies with a large spread in stellar mass tends to dilute the intrinsic dependence of CGM pressure on radius.  Two observational biases may also dilute the trend: (1) lines of sight at $r_{\rm proj} / v_c \ll 1$~Gyr pass through a greater range of CGM pressure and are more likely to intercept clouds with $P_{\rm CGM} \ll P_{\rm CGM} (r_{\rm proj})$, and (2) high-ionization clouds at $r_{\rm proj} / v_c \approx 1$~Gyr are more difficult to detect than their low-ionization counterparts.   Within the shaded area of the top panel, photoionization suppresses radiative cooling of the ambient medium. Precipitation model predictions in the shaded area therefore become increasingly invalid toward lower pressure.
\vspace*{1em}
\label{fig-P_vs_r}}
\end{figure*}
% ----------------------------------------

\section{Trends in the Data}
\label{sec-trends}

Figure~\ref{fig-U_vs_r} shows how those $U$ measurements depend on $M_*$ and projected distance from the galaxy's center ($r_{\rm proj}$).  No obvious dependence of cloud pressure on $r_{\rm proj}$ can been seen in the sample as a whole, but when the galaxies are grouped by stellar mass, each subset shows a decline in cloud pressure with radius.  Furthermore, gas pressure at a given radius in the highest-mass subset (COS-LRG) is approximately two orders of magnitude greater than in the three lowest-mass subsets.  Those trends still hold if the projected radius is divided by the circular velocity $v_c$ of an orbit in the gravitational potential of the galaxy's dark-matter halo, to give the dynamical timescale at $r = r_{\rm proj}$ (top panel of Figure~\ref{fig-P_vs_r}).  Dividing by $v_c$ is analogous to dividing by a virial radius but avoids introducing a potentially spurious cosmological dependence into the scaling.  The $v_c$ value for each galaxy comes from interpolating the $M_*$--$v_c$ relation of \citet{McGaugh+2010ApJ...708L..14M}, which is approximately $\log M_* \approx 10.9 + 4 \log (v_c / 200 \, {\rm km \, s^{-1}})$, with a scatter in $M_*$ of about 0.2~dex at fixed $v_c$, depending on assumptions about the mass-to-light ratio of the stellar population. If $n_{\rm H}$ in the photoionized clouds were simply proportional to the mean matter density enclosed within $r$, then $P_{\rm CGM}$ at $r_{\rm proj} / v_c$ would be nearly independent of galaxy mass.  The bottom panel of Figure~\ref{fig-P_vs_r} shows instead that $P_{\rm CGM} / v_c^4$ is much closer to being a mass-independent function of $r / v_c$, which is similar to the predictions of precipitation-limited CGM models (see \S \ref{sec-models}). 

A detrending analysis of the data demonstrates that the weighted sample variance of $P_{\rm CGM} / v_c^\zeta$ is minimized for $\zeta = 3.40 \pm 1.12$.  Before performing the minimization, we restrict the set of points included in the fit by removing all the points at at $r / v_c < 0.25$~Gyr, as well as all the clouds with higher-pressure companions along a given line of sight.  Doing so reduces but does not eliminate the influence of projection effects. ÊWe also remove all the points with predicted pressure $n_H T < 5 \, {\rm K \, cm^{-3}}$, because photoionization of the ambient medium is likely to alter radiative cooling and therefore the precipitation limit below that pressure \citep[e.g.,][]{Stern_2018arXiv180305446S}.  Fitting the remaining points to the formula
\begin{equation}
   P_{\rm CGM} = P_{\rm 300 \, Myr} 
                             \left( \frac {v_c} {\rm 200 \, km s^{-1}} \right)^\zeta
                             \left( \frac {r / v_c} {\rm 300 \, Myr} \right)^\alpha
\end{equation}
shows that the weighted sample variance is minimized at $\zeta \approx 3.4$ and $\alpha \approx - 1.7$, giving a weighted standard deviation $\approx 0.3$~dex. ÊWe therefore redo the fits with updated weighting obtained by adding a dispersion of 0.3 dex in quadrature to the observational uncertainties in $\log U$. Ê

% ----------------------------------------
\begin{figure}[t]
\begin{center}
\includegraphics[width=3.4in, trim = 0.1in 0.1in 0.0in 0.0in]{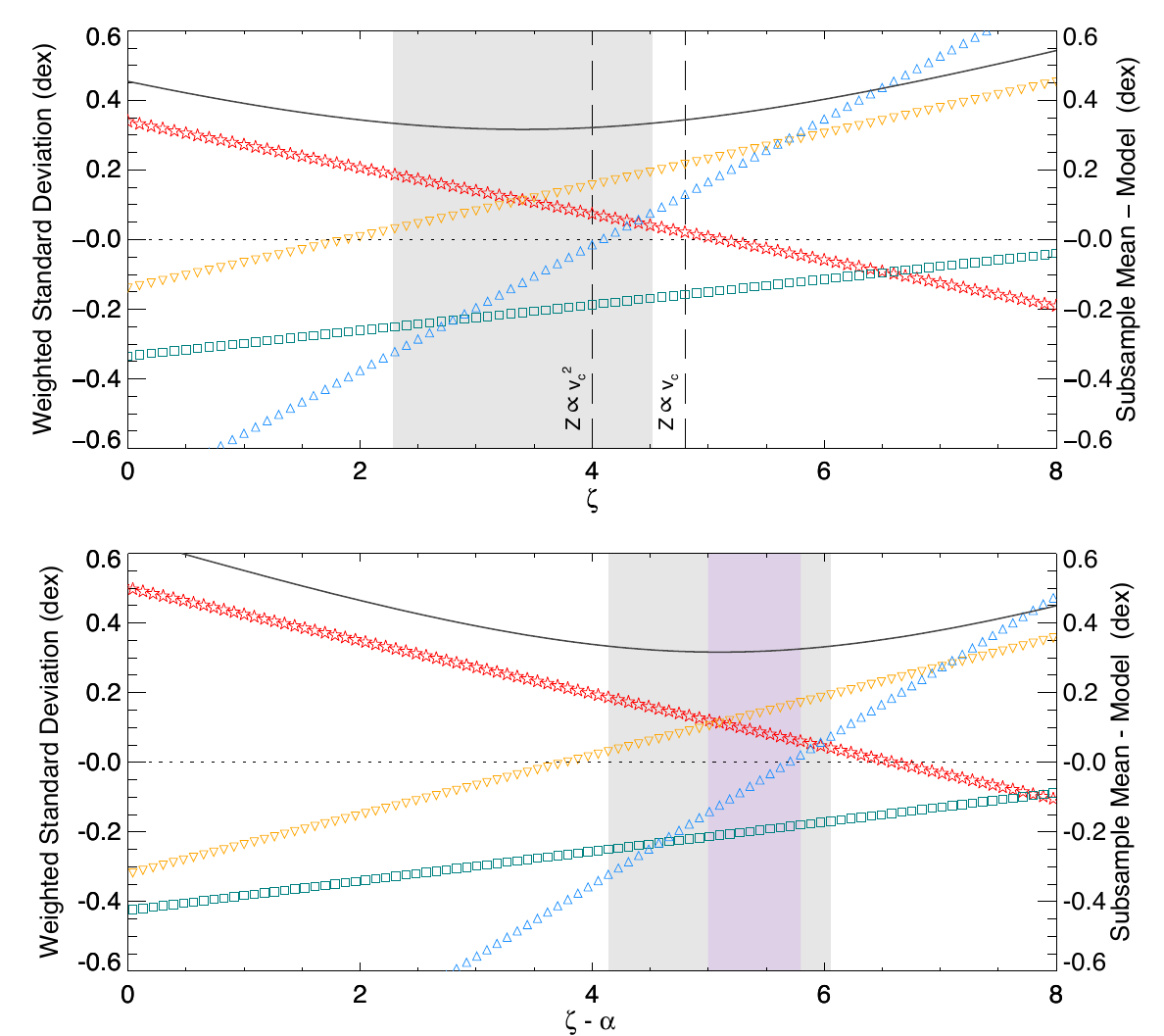} \\
\end{center}
\caption{ \footnotesize 
Parameters that minimize dispersion around the best fit to $P_{\rm CGM} \propto v_c^\zeta (r/v_c)^\alpha$.  Fits were made to a restricted data set that excludes points with either $r / v_c < 0.25$~Gyr or predicted $n_{\rm H} T < 5 \, {\rm K \, cm^{-3}}$. ÊÊCharcoal lines in each panel show the weighted standard deviation of $P_{\rm CGM} / v_c^\zeta$ from the best-fitting power-law function of $r / v_c$ (i.e. the value of $\alpha$ that minimizes the weighted sample variance).  Lines of symbols show how the offset of each weighted subsample mean deviates from the best-fitting power-law model, with symbol types corresponding to the same subsamples as in Figures~\ref{fig-U_vs_r}, \ref{fig-P_vs_r}, and \ref{fig-U_vs_UpNFW}.   The top panel shows that the weighted sample variance reaches a minimum value of 0.316 dex within the gray shaded region illustrating $\zeta = 3.40 \pm 1.12$, and vertical dashed lines show model predictions for CGM metallicity proportional to $v_c$ and $v_c^2$.   Gray shading in the bottom panel shows the $1\sigma$ region as a function of $\zeta - \alpha$, the $P_{\rm CGM}$--$v_c$ scaling parameter at fixed $r$, for which $\zeta - \alpha = 5.10 \pm 0.97$.   Purple shading in the bottom panel shows the range $5 \lesssim \zeta - \alpha \lesssim 5.8$ predicted by precipitation-limited models for galaxies in which $M_* \propto v_c^\beta$, with $4 \leq \beta \leq 5$ (see \S \ref{sec-models}). 
\vspace*{1em}
\label{fig-zeta}}
\end{figure}
% ----------------------------------------

The top panel of Figure~\ref{fig-zeta} shows the resulting dependence of weighted sample variance on $\zeta$. ÊIt is minimized at $\zeta = 3.40 \pm 1.12$ (max likelihood estimate) implying that $\zeta = 0$ is ruled out at greater than $3\sigma$ significance, assuming that the likelihood has a $\chi^2$ distribution. ÊA purely cosmological pressure profile would have $P_{\rm CGM} \propto v_c^2$  at fixed $r / v_c$, which is disfavored relative to $\zeta = 4$ but not ruled out.  ÊFigure~\ref{fig-zeta} also shows how deviations from the best fit depend on $\zeta$.  Each line of symbols corresponds to a subsample of the restricted data set.  For a given $\zeta$, the vertical position of a symbol represents the difference between the weighted mean pressure for that subsample and the best-fitting pressure profile for all the subsamples.  The slope of each line therefore depends on the mean $v_c$ for the galaxies in that subsample.  Red stars representing COS-LRG trend downward with increasing $\zeta$ because that is the highest-mass subsample.  The slopes of the other lines become increasingly more positive with decreasing mean mass.  They come closest together within the gray shaded region showing $\zeta = 3.40 \pm 1.12$.

The bottom panel of Figure~\ref{fig-zeta} shows the dependence of weighted sample variance on the parameter combination $\zeta - \alpha$, which specifies how $P_{\rm CGM}$ scales with $v_c$ at fixed radius. ÊIts minimum is at $\zeta - \alpha = 5.10 \pm 0.96$, implying that no dependence of $P_{\rm CGM}$ on $v_c$ is ruled out at greater than $5\sigma$ confidence. We obtain tighter constraints on $\zeta - \alpha$ than on $\zeta$ alone because of a covariance in the fit:  for small values of $\zeta$, a steeper pressure profile (larger absolute value of $\alpha$) brings the data points for low-mass systems closer to the best fit to the whole sample.

\section{Comparisons with Models}
\label{sec-models}

Models of precipitation-limited ambient media predict that $P_{\rm CGM} / v_c^4$ at $r / v_c$ should be nearly independent of galaxy mass, which is consistent with the detrending analysis.  This scaling results from the fundamental assumption that feedback keeps $\min ( t_{\rm cool} / t_{\rm ff} )$ roughly constant.  In this ratio, the cooling time is defined to be $t_{\rm cool} = 3nkT / 2 n_e n_{\rm H} \Lambda (T)$, where $n_e n_{\rm H} \Lambda (T,Z)$ is the radiative cooling rate per unit volume and $\Lambda (T,Z)$ applies to collisionally ionized gas at temperature $T$ with heavy-element abundances $Z$ times their solar values. ÊThe freefall time is defined to be $t_{\rm ff} = (2 r^2 / v_c^2)^{1/2}$. ÊWith those definitions, the corresponding upper limit on CGM pressure, expressed in terms of $n_{\rm H} T$, is
\begin{equation}
  P_{\rm CGM} = \frac {3kT^2} {10 \, \Lambda (T,Z)} 
                      \left( \frac {n} {2 n_e} \right) \left( \frac {v_c} {2^{1/2} r} \right)
 \label{eq-P_CGM}
  \; \; .
\end{equation}
This pressure limit scales as 
\begin{equation}
  \frac {P_{\rm CGM}} {v_c^4} \propto \Lambda^{-1}  \left( \frac {T} {v_c^2} \right)^2 \frac {v_c} {r}
  \label{eq-P_CGM_vc4}
  \; \; .
\end{equation}
because a volume-filling medium near hydrostatic equilibrium tends to have $kT \sim \mu m_p v_c^2$, where $\mu m_p$ is the mean mass per particle.  Dividing $P_{\rm CGM}$ by $v_c^4$ should therefore give a radial profile that depends primarily on the cooling function $\Lambda$, the dynamical time $r / v_c$, and the dimensionless ratio $kT/\mu m_p v_c^2$, which may vary with radius but should be largely independent of system mass.  

In precipitation-limited CGM models for galaxies in the mass range that we are considering, the cooling-function term in equation (\ref{eq-P_CGM_vc4}) is not expected to vary strongly with $v_c$, because its dependence on $Z$ tends to offset its dependence on $T$.  Fits to the cooling functions of \citet{sd93} give $\Lambda(T,Z)  \appropto (Z/T)^{0.8}$ for $10^{5.5} \, {\rm K} < T < 10^{6.5} \, {\rm K}$ and $0.2 < Z < 2$.  Consequently, the cooling-function term scales as $\Lambda \appropto v_c^{-0.8}$ for $Z \propto v_c$ and $\Lambda \sim {\rm const.}$ for $Z \propto v_c^2$.  Vertical dashed lines in Figure 3 show the resulting predictions for $\zeta$, which are close to the maximum-likelihood value for $Z \propto v_c^2$ and just outside the $1\sigma$ range for $Z \propto v_c$.

The {\em ambient} value of $Z$ is what determines the $P_{\rm CGM}$ prediction, and there are essentially no direct observational constraints on the scaling of ambient CGM metallicity with halo mass.  We will therefore outline the predictions that follow from the assumption that a galaxy's supernova ejecta are well mixed with the baryons associated with its halo, and compare them with observed metallicity trends among the stars and gas clouds within galaxies.  In that case, the CGM abundances in a galaxy population with $M_* \propto v_c^\beta$ should scale as $Z \propto M_* / M_{\rm halo} \propto v_c^{\beta - 3}$, giving $Z / T \propto v_c^{\beta - 5}$.  The $M_*$--$v_c$ relation of \citet{McGaugh+2010ApJ...708L..14M}, which has $\beta \approx 4$, then leads to $Z \propto v_c$ and $\Lambda \appropto v_c^{-0.8}$.  However, observations indicate that $\beta$ may become larger as $v_c$ declines.  For example, abundance matching of galaxies and halos \citep[e.g.][]{Moster_2010ApJ...710..903M} yields $M_* \appropto M_{\rm halo}^{5/3}$ for $9 \lesssim \log M_* \lesssim 10.5$, implying $\beta \approx 5$.  
%Additionally, the observed relationship between stellar mass and stellar metallicity among galaxies in the same mass range is $Z \appropto M_*^{0.4}$ \citep[e.g.][]{Gallazzi_2005MNRAS.362...41G}, which is consistent with both $\beta \approx 5$ and $Z  \propto v_c^2$ in the limit of efficient mixing.
Observational constraints on the relationship between stellar mass and gas-phase metallicity are harder to apply because of systematic uncertainties in the metallicity diagnostics.  For $9.5 \lesssim \log M_* \lesssim 10.5$, some analyses are consistent with $Z \appropto M_*^{0.4}$, and therefore with $\beta \approx 5$ \citep[e.g.,][]{Blanc_2019arXiv190402721B}, while others indicate a relationship closer to $Z \appropto M_*^{0.2}$ \citep[e.g.,][]{Sanchez_2017MNRAS.469.2121S}, which is more consistent with $\beta \approx 4$. And for $\log M_* \gtrsim 10.5$, both the stellar and gas-phase metallicities saturate near the solar value, indicating that a simple power-law model might not be adequate for expressing the relationship between $P_{\rm CGM}$ and $v_c$ over the entire range of stellar mass that we are considering.

On the other hand, assuming $M_* \propto v_c^\beta$ allows us to bring the model predictions one step closer to the data, because $v_c^4$ is really a proxy for $M_*$ in the detrending analysis of \S \ref{sec-trends}.  Fundamentally, the detrending analysis constrains $\zeta - \alpha$ in the power-law relation $P_{\rm CGM} \propto M_*^{(\zeta - \alpha)/\beta} r^{\alpha}$.  Assuming that $M_* \propto v_c^\beta$ and efficient mixing then converts equation (\ref{eq-P_CGM_vc4}) into the prediction 
\begin{equation}
  P_{\rm CGM}  \appropto {M_*^{(9  - 0.8 \beta)/\beta}} 
  	\left[ \left( \frac {T} {v_c^2} \right)^{2.8} \frac {1} {r} \right]
  \label{eq-P_CGM_M*}
  \; \; ,
\end{equation}
in which the factor in square brackets is presumed to depend only on $r$.  Within the range $4 \lesssim \beta \lesssim 5$, the model therefore predicts $5 \lesssim \zeta - \alpha \lesssim 5.8$, which lies within $1\sigma$ of the maximum-likelihood value derived from detrending the data (see Figure~\ref{fig-zeta}).  Extending the model down to $\beta \approx 3$, in which case $Z$ becomes independent of $v_c$, gives the prediction $\zeta - \alpha \approx 6.6$, which is more than 1$\sigma$ from the maximum-likelihood value but well within the $2\sigma$ range.  In other words, the observed dependence of $P_{\rm CGM}$ on $M_*$ and $r$ is much more consistent with precipitation-model predictions than with a CGM pressure that is independent of $M_*$ and $r$.

Accounting for the dependence of $P_{\rm CGM}$ on halo mass by assuming $\zeta = 4$ reveals that the radial pressure gradients predicted by precipitation-limited models are also similar to those found in the data.  The solid and dashed lines in Figures \ref{fig-U_vs_r} and \ref{fig-P_vs_r} show predicted pressure profiles from \citet{Voit_2018arXiv181104976V} for galaxies that would belong to each of the mass-segregated subsets.  In all the models, CGM pressure is determined by the precipitation-limit condition $\min (t_{\rm cool} / t_{\rm ff}) = 10$.  The resulting models have lower pressures at large radii than previously proposed models for hot atmospheres \citep{MoMiralda_1996ApJ...469..589M,MallerBullock_2004MNRAS.355..694M,Faerman_2017ApJ...835...52F} in which the minimum pressures are inconsistent with observations \citep{Werk_2014ApJ...792....8W}.
In each galaxy-mass subset, the data points generally track the corresponding model, but projection effects produce considerable scatter, for reasons best illustrated in Figure~\ref{fig-U_vs_r}.  Dotted lines in that figure connect measurements of clouds that all belong to a single galaxy but are slightly separated in velocity space.  According to the data, the spread in CGM pressure along at least some lines of sight can exceed two orders of magnitude.  We therefore take the highest-pressure cloud to be most representative of the CGM pressure at a distance $r \approx r_{\rm proj}$ from the galaxy, while recognizing that its pressure may still be just a lower bound on the maximum pressure along that sightline.

\begin{figure*}[t]
\begin{center}
\includegraphics[width=6.7in, trim = 0.1in 0.1in 0.0in 0.0in]{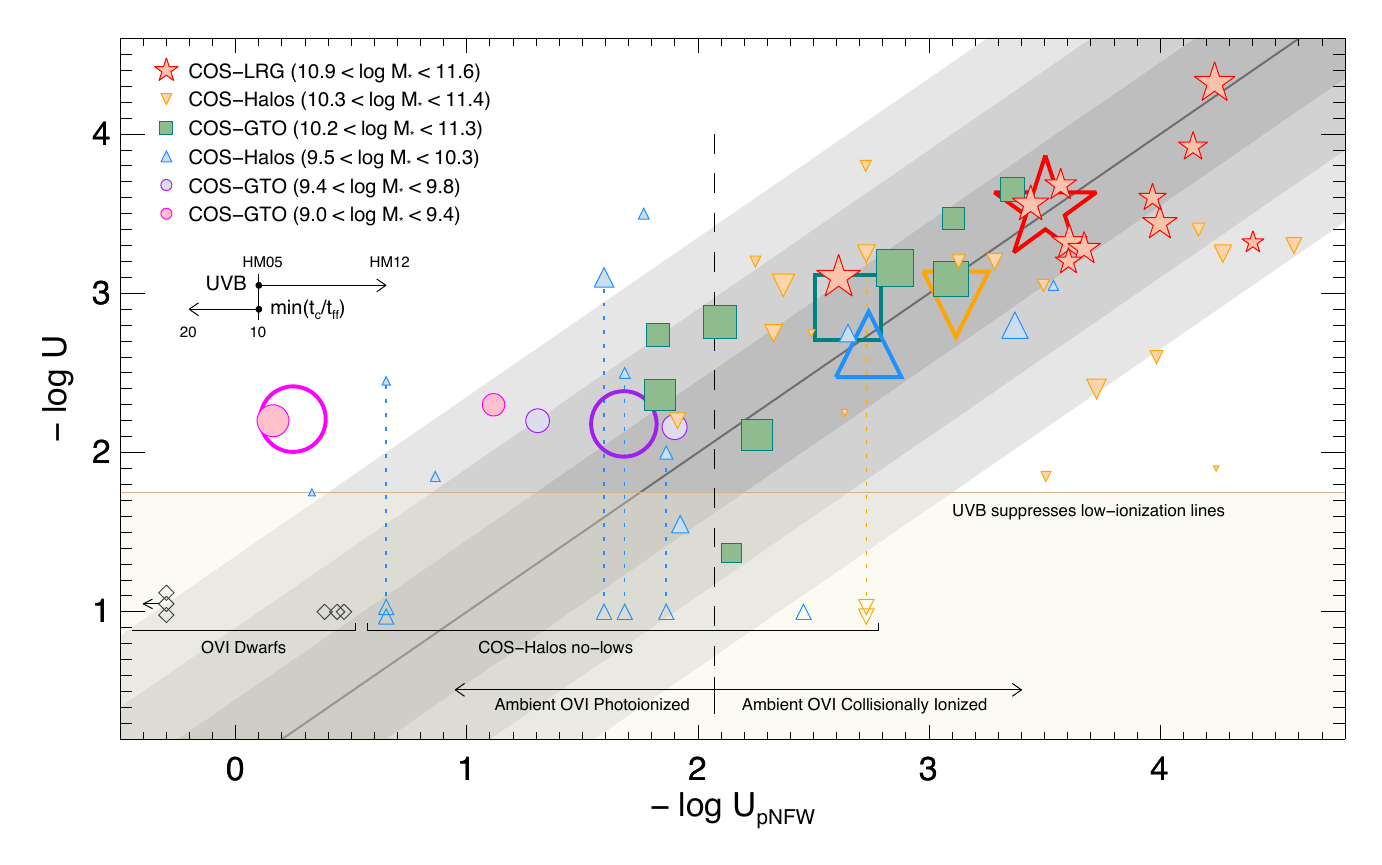} \\
\end{center}
\caption{ \footnotesize 
Comparison of predicted ionization level of CGM clouds ($U_{\rm pNFW}$) with observations.  Filled symbols have the same meanings as in Figure 2.  Large open symbols show weighted means for each of the subsamples.  Small open symbols show additional high-ionization systems in which $U$ is not well constrained.  The right-pointing arrow labeled UVB shows how the data points shift if the HM12 background is assumed instead of the HM05 background.  The left-pointing arrow labeled $ \min (t_{\rm cool} / t_{\rm ff})$ shows how the data points shift if $\min (t_{\rm cool} / t_{\rm ff}) = 20$ is assumed instead of $\min (t_{\rm cool} / t_{\rm ff}) = 10$ in the precipitation-limited (pNFW) models. A solid gray line shows where model predictions would match observations.  Shading around the line shows the expected $1\sigma$, $2\sigma$ and $3\sigma$ levels in the log-normal uncertainty distribution of $U_{\rm pNFW}$.  Ionization levels become more difficult to measure in the shaded high-ionization region below the horizontal tan line.  Small open triangles represent CGM clouds from the COS-Halos sample that produce O VI absorption lines but have no detectable low-ionization gas, called no-lows by \citet{Werk2016_ApJ...833...54W}.  They are plotted at $\log U = -1$ because their ionization levels can be no smaller than $U = 0.1$.  Open diamonds at the lower left represent similar O VI detections of the CGM around dwarf galaxies \citep{Johnson_OVIDwarfs_2017ApJ...850L..10J}.  Predictions for the three dwarf galaxies connected to the left-pointing arrow are smaller than the lower limit of the plot. The vertical dashed line corresponds to $n_{\rm H} T \approx 5 \, {\rm K \, cm^{-3}}$.  Toward the left of that line, photoionization increasingly suppresses radiative cooling of the ambient medium, allowing CGM pressures to be greater than the pNFW models predict.
\vspace*{1em}
\label{fig-U_vs_UpNFW}}
\end{figure*}

Figure~\ref{fig-U_vs_UpNFW} shows how the ionization level predicted for a cloud at radius $r_{\rm proj}$ by a precipitation-limited model for a galaxy with stellar mass $M_*$ compares with observations.  The predicted ionization levels ($U_{\rm pNFW}$) are determined by a model for the UV background and a pressure calculated according to the pNFW prescription from \citet{Voit_2018arXiv181104976V}. ÊEach pressure calculation depends on projected radius, the maximum circular velocity $v_c$ of the gravitational potential, and an assumption about the element abundance $Z$ in the ambient CGM. ÊFor the COS-LRG galaxies, we used ``Zgrad" models, which assume solar abundances at small radii and a gradual decline to 0.3 solar at large radii. ÊFor the higher-mass COS-Halos and COS-GTO galaxies, we used models with 0.3 times solar abundances. ÊFor the lower-mass COS-Halos and COS-GTO galaxies, we used models with 0.1 times solar abundances.  (Note that abundances in the ambient CGM are not necessarily identical to those observed in the cooler photoionized clouds, because at least some of those clouds may not have condensed out of the ambient gas.)  A dark gray line indicates where observations would match those predictions.  Uncertainties in mapping $M_*$ onto a particular CGM model add considerable horizontal scatter, which is approximately represented by the shaded region around the line, and the individual data points generally follow the model predictions, with a scatter similar to the expected dispersion.  Large open symbols corresponding to each mass-segregated subsample represent weighted subsample means and generally lie along the dark gray line, with no significant dependence on galaxy mass, except perhaps at $\log M_* \approx  9.0$.

% ----------------------------------------
% ----------------------------------------

\section{Concluding Thoughts}

These results indicate that galaxies in the stellar mass range $9 < \log M_* < 11.5$ adhere to the same regulating principle that governs the CGM around higher-mass galaxies, in that the cooling time of volume-filling ambient gas cannot fall much below $10 t_{\rm ff}$ at all radii. In high-mass galaxies, the feedback that limits cooling comes from black hole accretion.  In low-mass galaxies it comes mostly from supernovae, but a role for black holes cannot be ruled out.  In this marginally unstable state, condensation of ambient gas may be producing at least some of the photoionized $10^4$~K clouds that are embedded within it, and the predicted condensation rate is similar to a galaxy's time-averaged rate of star formation \citep{Voit_PrecipReg_2015ApJ...808L..30V}.  Around dwarf galaxies of even lower mass ($\log M_* < 9$), photoelectric heating suppresses radiative cooling of ambient circumgalactic gas at the predicted pressure \citep{Stern_2018arXiv180305446S}, with implications for precipitation that have yet to be modeled.  However, it is likely that the UV background then maintains the CGM in a state that maximizes the abundance of the O$^{5+}$ ion and allows those dwarf galaxies to produce strong O VI absorption lines \citep{Johnson_OVIDwarfs_2017ApJ...850L..10J}.

\vspace*{2em}

G.M.V. acknowledges helpful conversations with Gus Evrard and has been supported in part by {\em Chandra} Science Center grant TM8-19006X.  H.W.C. and F.S.Z. acknowledge partial support from HST- GO-14145.01A and NSF AST-1715692 grants. J.K.W. acknowledges support from a 2018 Alfred P. Sloan Research Fellowship and NSF-AST-1812531. B.W.O. was supported in part by NSF grants PHY-1430152, AST-1514700, OAC-1835213, by NASA grants NNX12AC98G, NNX15AP39G, and by HST AR \#14315. G.L.B. was partially supported by NSF grant AST-1615955 and NASA grant NNX15AB20G.

%\newpage

\bibliographystyle{apj}
%\bibliography{precipitation}

\end{document}